# Quantifying the uncertainty of molecular dynamics simulations : Good-Turing statistics revisited

Vasiliki Tsampazi & Nicholas M. Glykos*

Department of Molecular Biology and Genetics, Democritus University
of Thrace, University campus, 68100 Alexandroupolis, Greece, Tel +30-25510-30620,
Fax +30-25510-30620, https://utopia.duth.gr/glykos/ , glykos@mbg.duth.gr

# Abstract

We have previously shown that Good-Turing statistics can be applied to molecular dynamics trajectories to estimate the probability of observing completely new (thus far unobserved) biomolecular structures, and showed that the method is stable, dependable and its predictions verifiable. The major problem with that initial algorithm was the requirement for calculating and storing in memory the two-dimensional RMSD matrix of the currently available trajectory. This requirement precluded the application of the method to very long simulations. Here we describe a new variant of the Good-Turing algorithm whose memory requirements scale linearly with the number of structures in the trajectory, making it suitable even for extremely long simulations. We show that the new method gives essentially identical results with the older implementation, and present results obtained from trajectories containing up to 22 million structures. A computer program implementing the new algorithm is available from standard repositories.

# 1 Introduction

Biomolecular systems are so complex that it is practically impossible for a simulation to sample all feasible structures of a macromolecule. In most cases, however, faithful sampling of all accessible structures is *not* needed, and attempting to do so would be a waste of computational resources. For example, if the question that we wish to answer —given a trajectory— is whether a protein structure is stable at a given temperature, then what we need is an estimate of the probability of observing significantly different structures if the simulation was continued. If we could somehow establish, for example, that the probability of observing a structure that differs by more than 1.0Å RMSD from those already observed is less than $10^{-9}$, then we could safely conclude that the structure is indeed stable, and that continuing the simulation is unnecessary. What is needed, is a method which based on the information contained in an existing simulation, would estimate the probability of observing completely new (thus far unobserved) structures. We have previously described such a method[1], which is based on the application of Good-Turing statistics[2,3] to molecular dynamics trajectories and showed that the method is stable, dependable and its predictions verifiable.

We will introduce the subject of using Good-Turing statistics for estimating the uncertainty of molecular dynamics simulations starting from the end, ie. by first presenting the form of the results produced by this method[1].

Fig.1 shows the end-product of the Good-Turing analysis in the form of the results obtained from four independent molecular dynamics trajectories of different proteins and peptides. These four simulations cover the whole range from very stable *NpT* simulations performed in folded state (ROP protein[4]), to extensive folding simulations of two peptides (CLN025[5], 6NM2[6]) and one mini-protein (FipWW domain[7]), and cover simulation lengths ranging from 6.6 to 20 μs. The following paragraphs discuss the form of these graphs, the information content they carry, and their application in quantifying the uncertainty of molecular dynamics simulations.

What the Good-Turing method does is to estimate —based on a given trajectory— the probability *(P)* of observing a new structure that differs by more than *(x)* Å RMSD from all structures that we already observed in the given trajectory. All curves start with very high probability values for low RMSDs which is equivalent to saying that "it is very probable that if you extend the simulation you will observe structures that are very similar to those you already observed". But careful examination of Fig.1 shows that even in this low RMSD part of the diagram, the curves differ significantly between different simulations. For example, the folding simulations of the CLN025 & 6NM2 peptides start at an RMSD of ~0.2Å, whereas ROP and FipWW start significantly later at RMSDs of ~0.5Å or more. Even this seemingly small difference is, however, meaningful : the two peptides are very short, of the order of ten residues, whereas ROP, for example, is a 126-residue homodimer. The differences in the low RMSD part of the graphs signify what we already expected, ie. that it is far more probable to observe very low RMSDs when you have a small number of residues, than it is when you study a reasonably large protein.

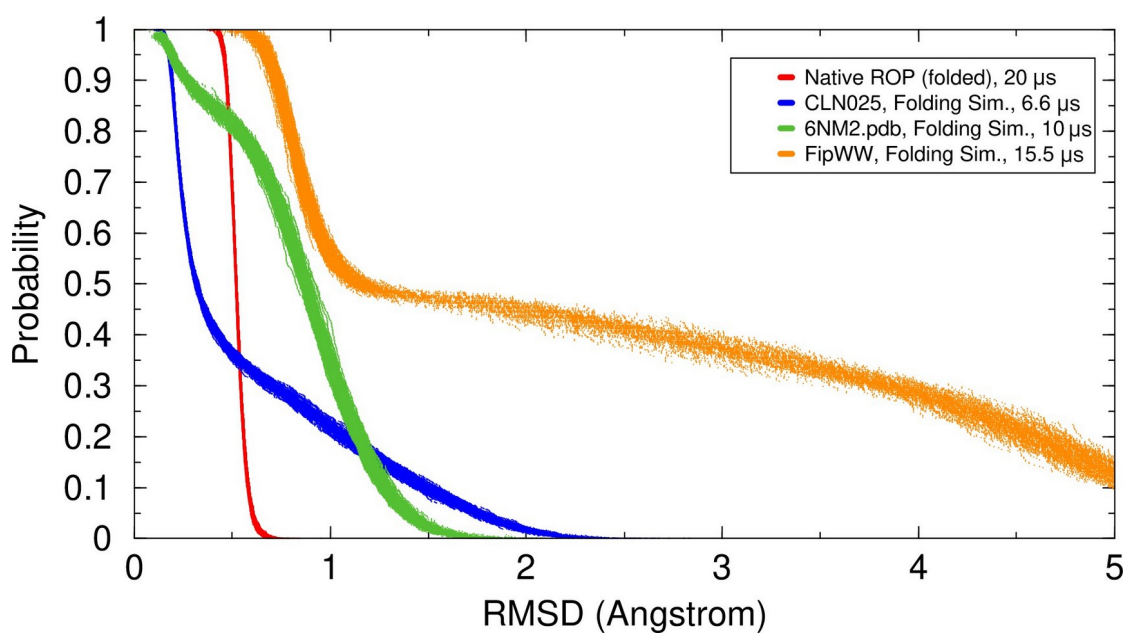

**Fig.1** Good-Turing probability curves of four different molecular dynamics trajectories. See text for an extensive discussion of this Figure.

The exact form of the graphs and how (and at what RMSD values) they approach small probabilities is a direct quantification of the structural uncertainty that is left unaccounted for by the already recorded trajectories. For example, the differences between the ROP and FipWW curves clearly indicate that for the simulation of ROP (which was performed in the folded state) very little uncertainty remains unaccounted for by the current trajectory, whereas the curve for the FipWW folding simulation leaves little doubt that completely different structures will be observed if the simulation is extended. It is worth noting that the method as it stands now, can also provide an estimate for the answer to the following question : "If we extend the simulation by *doubling* the simulation time, what RMSD should we expect for the most different newly observed structure ?". To continue with the current example, for the ROP simulation the method estimates that if we double the simulation time (from 20μs to 40μs), then the most different structure that we should expect to observe would differ by no more than approximately 0.95Å RMSD from any of the already observed structures (in the first 20μs). In contrast, for the folding simulation of the FipWW protein, the method estimates that if we double the simulation time (from 15.5μs to 31μs), then the most different structure that we should expect to observe may differ by as much as 10.98Å RMSD from any of the already observed structures. We believe that this expected RMSD of ~11Å for a mini-protein with only 34-residues satisfactorily captures the enormous configurational space available to folding simulations.

As a last example of the information content of the Good-Turing probability curves, we will discuss and compare the results obtained from the folding simulations of the CLN025 and 6NM2 peptides (blue and green curves in Fig.1). These two peptides are comparable in length (10 and 9 residues respectively), both simulations were performed using adaptive tempering to enhance sampling, and the timescales of the two trajectories are comparable (at 6.6 and 10 μs respectively). Nonetheless, the Good-Turing results differ significantly : the CLN025 peptide appears to have better sampling at low RMSD values, but worse (higher probabilities) for larger RMSDs which creates a cross-over point at approximately 1.2Å. The explanation for this behavior lies in the pronounced differences of the folding landscapes and structural stabilities of the two peptides as shown in Fig.2.

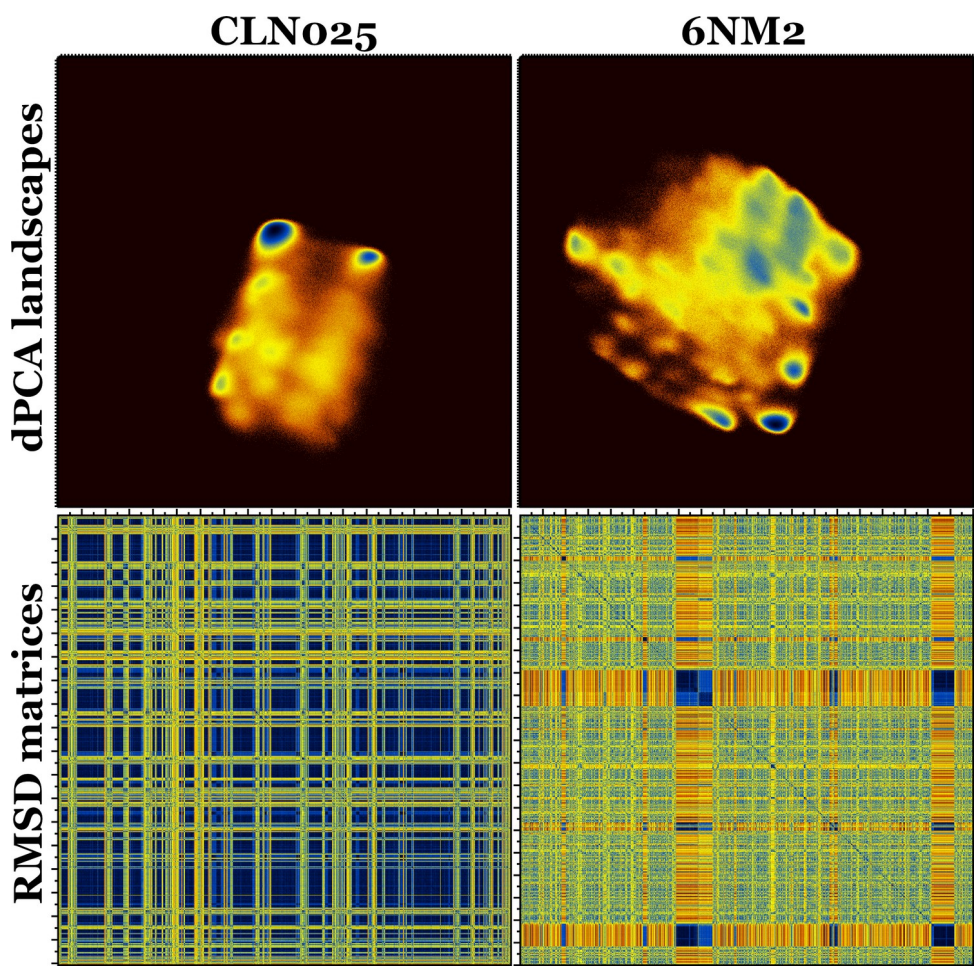

**Fig.2** Dihedral PCA-derived folding landscapes (upper row) and RMSD matrices (lower row) for the CLN025 and 6NM2 peptides. The dPCA-derived landscapes are the log density distributions of the projection of the trajectories on their respective top two principal components (high density → dark blue). For the RMSD matrices the origin is at the upper left-hand side corner, and the color coding ranges from dark blue (very low RMSDs, peptides stably folded), through yellow (medium RMSDs), to dark red (large RMSDs, peptides unfolded).

CLN025 is a very fast and stable folder[5], which spends most of the simulation time visiting repeatedly its native β-hairpin structure. This can be seen from its corresponding RMSD matrix in Fig.2 which is characterized by the overabundance of very low RMSD values (corresponding to blue colors in these diagrams). Unfolding events (yellow-red colors) are short lived and quickly lead to re-folding events. This over-stabilization of the native structure is brought forward by the dihedral PCA-derived folding landscape[8–12] which is essentially a single minimum landscape (remembering that the folding landscapes are on a logarithmic scale).

6NM2 on the other hand, is a flexible and mostly disordered peptide . It does have a more stable helical conformation which is visited twice during the simulation (the prominent blue-colored areas in the lower panel of Fig.2), but on the whole the peptide spends most of its simulation time visiting new conformations. In the light on this analysis, the predictions made by the Good-Turing probability curves are meaningful. For example, the higher probabilities at very low RMSDs for the CLN025 peptide encode the evidence obtained from its simulation, ie. that the peptide is so stable that it is very likely to repeatedly visit its native structure (which will have very low RMSD from the structures already observed). 6NM2 on the other hand, is highly flexible and is rightly expected to visit many new different structures, leading to the translation of the curve to higher RMSDs. The differences between the two curves at higher RMSDs are also consistent with our expectations : 6NM2 by being mostly disordered, has better sampled its configurational space, considering also the longer simulation time, and that it is a shorter peptide compared to CLN025. The CLN025 peptide on the other hand, by being mostly folded, has not extensively explored its folding landscape which leads to significant probabilities for higher RMSDs.

In the sections that follow we present the technical aspects of the calculations involved, discuss the practical application of the computer program that we distribute, and close by an extensive discussion of its perceived limitations.

## 2  Algorithm

The essential result of the Good-Turing frequency estimation that we need for the application to molecular dynamics trajectories is the following : given a pool of $N$ biomolecular structures that we observed in a trajectory, the probability $P$ of observing a thus far unobserved 'species' (i.e. a thus far unobserved structure) is given by

$$P = N_1 / N$$

where $N_1$ is the number of distinct molecular conformations for which only one individual structure was observed. This presentation immediately illustrates the major problem with the application of the method : molecular dynamics trajectories are essentially continuous in the configurational space, and are not organized in the form of distinct 'species'. The important contribution[1] was to realize that what is needed is a sub-sampling of the original trajectory in such a way that successive structures (in this sub-sampled trajectory) are not mechanistically correlated (due to the short time interval used for recording structures in the original trajectory) and can be, thus, be treated as "distinct species from a distribution containing a currently unknown number of species". The procedure for determining this sub-sampling factor from a molecular dynamics trajectory is discussed in the next section.

Once the required sub-sampling factor is known, the procedure as originally described is complete :

1. The sub-sampled RMSD matrix is extracted.
2. This sub-sampled-RMSD matrix is treated as a distance matrix, and a dendrogram is constructed using established hierarchical clustering methods.
3. The tree is cut at successively higher RMSD levels, and for each such cut the number of clusters with just one member ($N_1$) is calculated, as well as the corresponding probability $P = N_1/N$.
4. The final probability curve of *P* vs. RMSD is calculated.

This procedure was shown to work well, and to give dependable, consistent and verifiable results[1]. The major problem with its application, however, is the very significant memory requirements arising from the need to construct the full two-dimensional RMSD matrix of the trajectory. We believe that the need to calculate, store, and analyze the 2D RMSD matrix can be altogether avoided as follows.

Imagine examining an isolated row (say, the 1000th row) taken from the 2D RMSD matrix. This row contains all RMSD values between the corresponding (1000th) structure and all other structures recorded in the trajectory. The maximum of these RMSD values (ie. the maximum of the corresponding row from the matrix) is the answer to the following question : "Using as a reference structure the 1000th structure recorded, find the value of the RMSD for the structure that differs the most from this reference". If we now calculate the maximal RMSDs from *all* rows of the matrix and sort them numerically, then the largest value in this sorted list will correspond to the maximum RMSD observed in the *whole* 2D matrix, and will thus correspond to the one structure that differs the most from all the other structures observed. Imagine now that we had somehow calculated a dendrogram from the RMSD matrix and produced a tree (as described in steps 2 & 3 above). Then, cutting this tree at this (max-of-max) RMSD we found, would give one and only one cluster with just one member for the given RMSD value (corresponding to the one structure that differs the most from all the other structures). Clearly, this RMSD should be associated with a probability of $P = 1/N$ which gives us the first point of the ($P$ vs. RMSD) curve. The second largest RMSD from the sorted list will correspond to the next bifurcation point of the (binary) tree, and thus the second largest RMSD should be associated with a probability of $P = 2/N$. Continuing this procedure we associate each of the per-row maximal RMSDs with its corresponding probability value and we thus construct the sought ($P$ vs. RMSD) curve. The important thing to notice here, is that in order to calculate this list of maximal RMSDs (one per line), we do *not* need to store and process the whole matrix. We only need to calculate isolated lines from the matrix (one at a time) and only store one number (the maximum RMSD) from each line. In other words, we trade speed of execution for physical memory requirements.

To summarize, the new algorithm that altogether avoids the calculation of the whole 2D matrix is the following :

1. Determine the sub-sampling factor ($s$) needed. See next section for how this is achieved.

2. For each structure in the trajectory with a step of ($s$), use this structure as reference, calculate the RMSDs between this structure and all other recorded structures, and from all those RMSDs only keep and store the maximum RMSD observed.

3. Sort in descending order this list of ($N$) maximal RMSDs.

4. Successively associate each of these maximal RMSDs with a probability of $P = i/N$ where ($i$) is the one-based index of the RMSD under examination in the sorted list.

5. Emit the final probability curve of $P$ vs. RMSD.

Depending on the sampling factor ($s$), this procedure is repeated for different initial offsets within the interval defined by *(s)*. The result is not a single curve, but a family of curves (as shown in Fig.1) whose variance provides an estimate of the uncertainty in the probability estimation procedure.

Fig.3 shows a direct comparison between the Good-Turing curves obtained from the old and the new implementations for the case of two independent trajectories (the ROP and CLN025 simulations shown in Fig.1). The excellent agreement between the two methods is notable especially when considering that one method is based on constructing the 2D matrix and perform hierarchical clustering, whereas the second only involves a sorting of the maximum-per-line RMSDs. Based on our calculations with tens of trajectories[4–7,13–20], this level of agreement is not always attainable, but the reason for that probably lies with the different methods used for determining the sub-sampling factor *(s)* discussed in the next section.

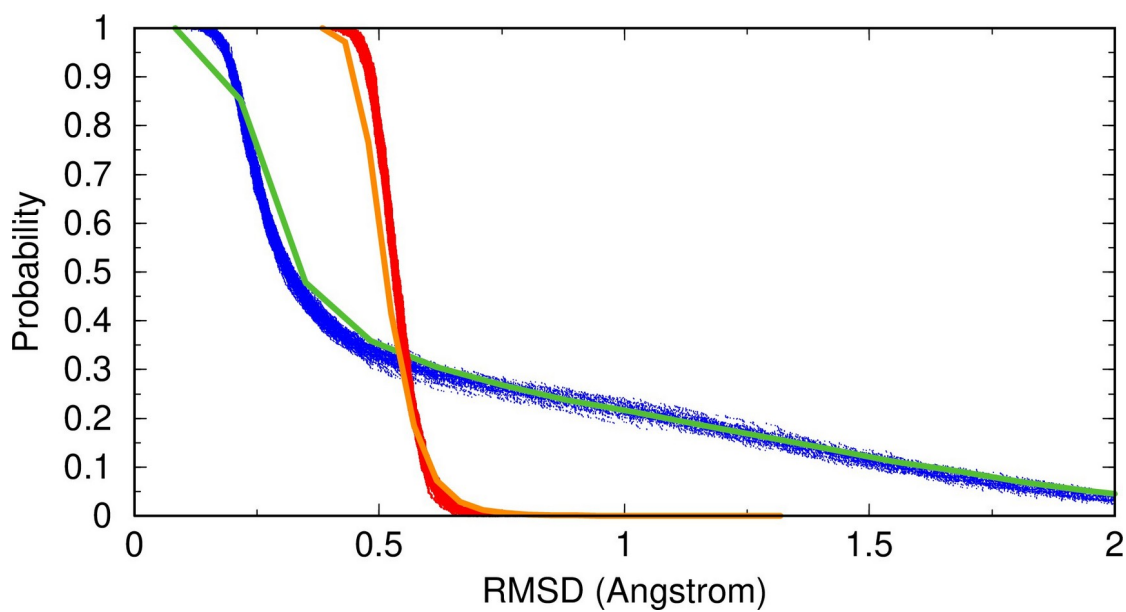

**Fig.3** Comparison of the results obtained from classical Good-Turing (green and orange curves) versus the results obtained from the new algorithm (blue and red scatter plots) for two independent trajectories. See text for details.

# 3   Implementation

The algorithm described in the previous section is so simple and straightforward that no further elaboration of its implementation details is required. What must be discussed, however, is the procedure needed for estimating the value of the sub-sampling factor (*s*) discussed above. This factor is the step/stride (along the original trajectory) needed to guarantee that successive structures are not mechanistically correlated due to the very short time interval between them. The principal idea for estimating (*s*) is that it will correspond to a time interval ($\delta t$) so long that the *maximal* RMSDs observed between all possible pairs of structures separated by ($\delta t$) stop increasing and converge to a stable plateau. The algorithm is the following :

1. Starting from a small value of a time interval ($\delta t$), calculate all possible values of RMSD between all possible pairs of structures separated by this time interval. Note that these RMSDs are located on a superdiagonal of the RMSD matrix (the one corresponding to the chosen value of $\delta t$).
2. Sort these RMSD values, and only keep the largest.
3. Repeat the procedure for different offsets within the given interval (to obtain a mean and a standard deviation for the given $\delta t$).
4. Repeat the procedure for increasing values of ($\delta t$).
5. In the graph of ($\delta t$ vs. RMSD) locate the point (*s*) at which the RMSDs reached a plateau and converged.

Fig.4 shows the form of the primary results obtained from this procedure for the case of the 6NM2 peptide discussed above (note that this Figure shows the raw data *before* attempting to determine the value of *s*). This diagram clearly illustrates why the determination of the sampling factor (*s*) is probably the single most difficult and weak step of our algorithm : In the presence of noise and of different time scales, determining the point at which the RMSDs converge not only is not straightforward, it may even be questionable. For a solid example, examination of Fig.4 shows that there are two points where the RMSDs reach an apparent plateau, the first at the ~110th superdiagonal, the second at the ~230th superdiagonal. Although in this example we can easily discern the general trend (and correctly conclude that the second solution is the correct one), encoding this algorithmically leads to solutions that are sensitive to noise.

In the previous version of our procedure[1] we tackled this problem by fitting the raw data to a generalized limiting diode equation (black line in Fig.4) and then attempting to identify points that deviate significantly from the curve (towards higher RMSDs). In this version of the algorithm we have resorted to using a different procedure which we believe is more stable and predictable.

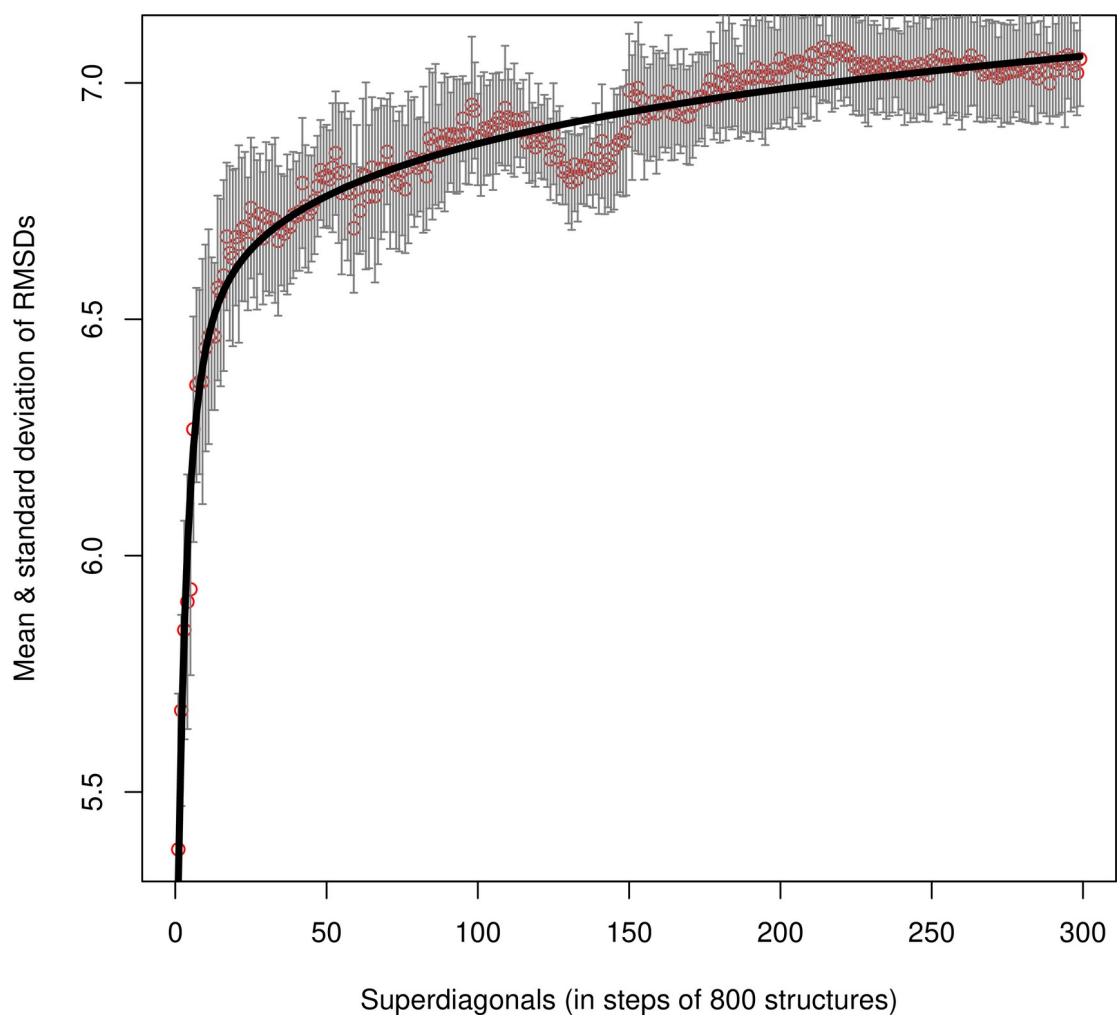

**Fig.4** Determination of the sampling factor : Raw data. This graph shows the primary data obtained from a folding simulation of a peptide. The horizontal axis corresponds to successive superdiagonals of the RMSD matrix (and, thus, corresponds to increasing time intervals $\delta t$), the vertical axis are the maximal RMSDs observed for the given ($\delta t$). Because for each ($\delta t$) different offsets within the interval are being tested, the result is not single RMSD value, but a distribution which is characterized by a mean and a standard deviation as depicted here. The black line is the non-linear fit of a generalized limiting diode equation, see text for details.

The principal idea with the new method we distribute is based on a piecewise linear segmentation of the primary data, followed by a heuristic search of the longest segment that fulfills predetermined criteria of length and slope. The piecewise linear segmentation is performed with the *R* package *dpseg()*[21,22], and Fig.5 shows an example of how the primary data (shown in Fig.4) would have been interpreted. In the case examined here, our heuristic implementation would have correctly selected the beginning of the last (blue) segment as the sampling factor for the Good-Turing analysis.

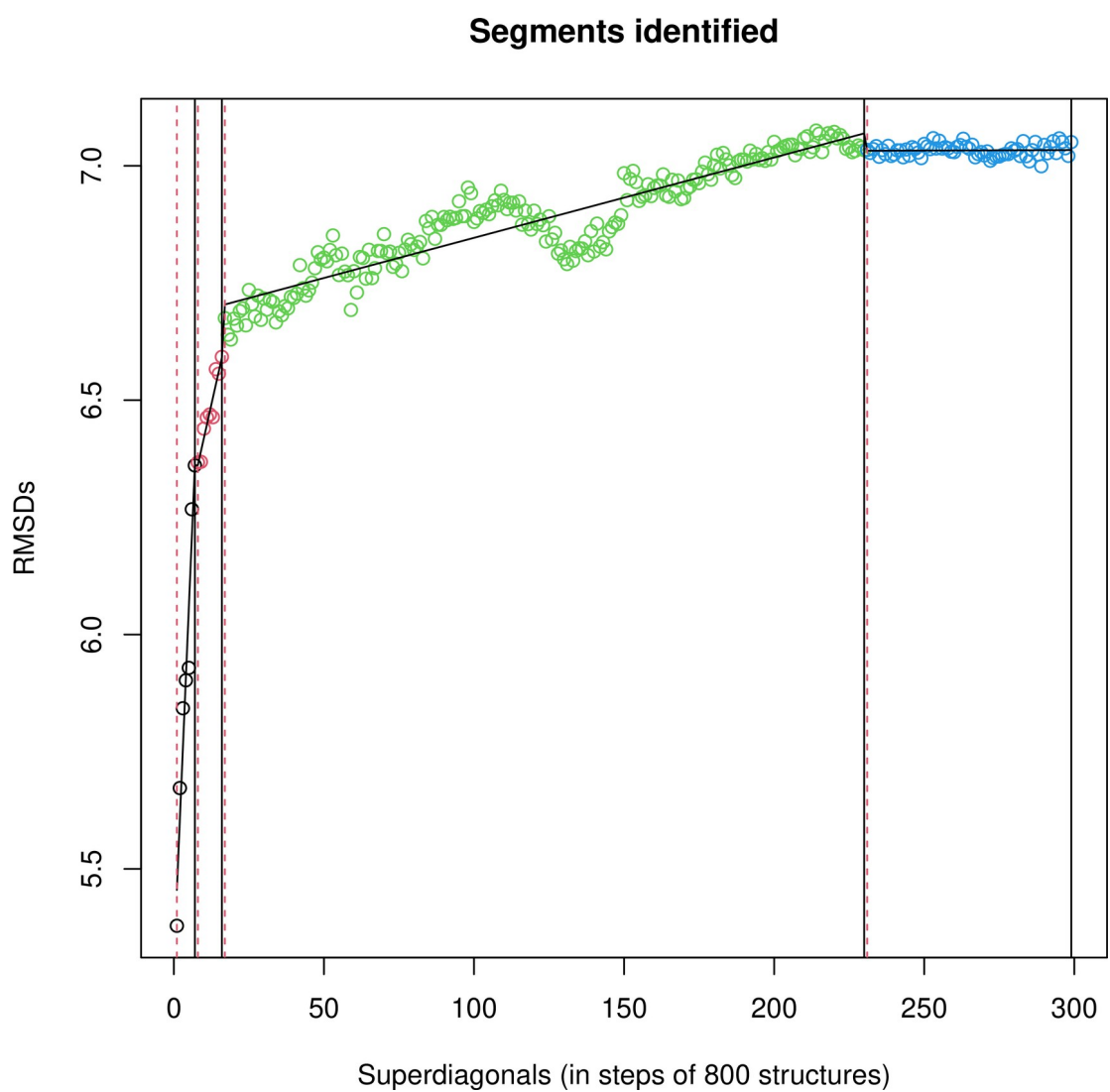

**Fig.5** Determination of the sampling factor : Linear segmentation. This graph shows an example of how the primary data (identical with those shown in Fig.4) can be interpreted in terms of successive linear segments obtained through the dynamic programming approach encoded in the *R* package *dpseg()*, see text for details. In this example, our algorithm would have picked a sampling factor corresponding to the beginning of the last (blue) segment.

# 4   Discussion

We are convinced that the probabilistic Good-Turing estimates such as those shown in the diagrams of Fig.1 are a useful addition to the established set of procedures[23] aiming to quantify the uncertainty of molecular dynamics trajectories of biological macromolecules. The method appears to be dependable, and the results are meaningful and verifiable. Having said that, we must emphatically note that there is a fundamental weakness hidden in our procedure that can not be bypassed algorithmically : Good-Turing statistics are strictly valid only for the case of sampling *distinct* objects from a pool containing an unknown number of such objects. By selecting a single sampling factor ($s$) for our analysis, we are essentially selecting a specific *timescale* of the structural events that we wish to analyze. The fact that we have not detected artifacts in our tests with a multitude of trajectories, is probably the result of how analysts design simulations. For example, if the aim of our analysis is to quantify the uncertainty of the conformations of the amino acid residues located in the active site of an enzyme, then it is highly unlikely that we would have elected to perform the simulation at a temperature where the whole enzyme may unfold within the timescale of the simulation. To put this differently, we have not observed artifacts from the mixing of different timescales not because they are not present, but because we actively design our simulations in such a way as to examine and analyze predetermined timescales that are relevant for the question in hand. Additionally, and because our algorithm will by default select the longest timescale consistent with the structural changes observed, it also works well with folding simulations (which necessarily contain all shorter timescales), with the silent assumption being that when you do a folding simulation, you do not care about structural details such as, for example, the stability of the conformations of a given tryptophan residue.

This discussion about the mixing of different timescales in simulations also explains the difficulties with the technically weakest part of the program that we distribute, namely the determination of the sampling factor ($s$) as discussed in section §3. We believe that the crux of the matter is that when well-separated timescales do mix in the same trajectory, then what we are trying to interpret in terms of a 'growth' curve with a single plateau, may in reality contain multiple plateaus corresponding to the different timescales present. As already mentioned, the current version of our algorithm will by default use the longest timescale observed (which is almost always the safest course of action since it avoids underestimating the structural uncertainty that remains unaccounted for by the existing simulation).

To conclude, the application of Good-Turing statistics for quantifying the uncertainty of molecular dynamics simulations appears to be a powerful technique. Being able to quantitatively answer questions like "Will doubling the simulation time from 20μs to 40μs significantly affect my conclusions ?" is to our mind a significant step towards solidifying the validity of the conclusions drawn from molecular dynamics simulations.

## Software Availability

All calculations reported in this communication were performed with free open source software which is immediately available for download *via* https://github.com/glykos/GoodTuring and https://github.com/glykos/carma.